\documentclass[reprint,twocolumn,superscriptaddress,secnumarabic,amssymb, longbibliography,nobibnotes, aps,pra]{revtex4-1}
\usepackage[T1]{fontenc}
\usepackage[utf8]{inputenc}
\usepackage{amssymb}
\usepackage{amsmath}
\usepackage{color}
\usepackage{xcolor}
\usepackage{hyperref}
\usepackage{verbatim}
\usepackage{graphicx}
\usepackage{gensymb}
\usepackage{comment}
\usepackage{ulem}

\begin{document}

\preprint{APS/123-QED}

\title{Probing La-based nickelates with Ni 1$s$ core-level photoelectron spectroscopy}

\author{Daisuke Takegami}
\email[]{dtakegami@tmu.ac.jp}
\affiliation{Department of Physics, Tokyo Metropolitan University, Hachioji, 192-0397, Japan}
\affiliation{Department of Applied Physics, Waseda University, 3-4-1 Okubo, Shinjuku-ku, Tokyo 169-8555, Japan}
\affiliation{Max Planck Institute for Chemical Physics of Solids, N{\"o}thnitzer Str. 40, 01187 Dresden, Germany}
%%%%%%%%%%%%%%%%%%%%%

%%%%%%%%%%%%%%%%%%%%%
\author{Naoki Ito}
\affiliation{Department of Physics and Electronics, Osaka Metropolitan University, 1-1 Gakuen-cho, Nakaku, Sakai, Osaka 599-8531, Japan}
%%%%%%%%%%%%%%%%%%%%%

%%%%% Experiment: HAXPES Beamtime July 2024 %%%%%
\author{Koto Fujinuma}
\affiliation{Department of Applied Physics, Waseda University, 3-4-1 Okubo, Shinjuku-ku, Tokyo 169-8555, Japan}

\author{Masato Yoshimura}  %yoshimur@spring8.or.jp
\affiliation{National Synchrotron Radiation Research Center, 101 Hsin-Ann Road, 30076 Hsinchu, Taiwan}
%%%%%%%%%%%%%%%%%%%%%%%%%%%%%%%%%%%%%%%%%%%%%%%%%%%%%%%%%
%%%%   Uncomment relevant ones    %%%
%%%%%%%%%%%%%%%%%%%%%%%%%%%%%%%%%%%%%%%%%%%%%%%%%%%%%%%%%

%\author{Edgar Abarca Morales}
%\affiliation{Max Planck Institute for Chemical Physics of Solids, N{\"o}thnitzer Str. 40, 01187 Dresden, Germany}

%\author{Mizuki Furo}
%\affiliation{Department of Physics and Electronics, Osaka Metropolitan University, 1-1 Gakuen-cho, Nakaku, Sakai, Osaka 599-8531, Japan}

% \author{\color{red}Synthesis People}
% \affiliation{Somewhere}

%%%%%%%%%%%%%%%%%%%%%
% Sample synthesis - Films,
\author{Grace A. Pan}
\affiliation{Department of Physics, Harvard University, Cambridge, 02138 USA}

\author{Dan Ferenc Segedin}
\affiliation{Department of Physics, Harvard University, Cambridge, 02138 USA}

\author{Qi Song}
\affiliation{Department of Physics, Harvard University, Cambridge, 02138 USA}

\author{Hanjong Paik}
\affiliation{Platform for the Accelerated Realization, Analysis and Discovery of Interface Materials (PARADIM), Cornell University, Ithaca, 14853 USA}
\affiliation{School of Electrical and Computer Engineering, University of Oklahoma, Norman, OK 73019, USA}
\affiliation{Center for Quantum Research and Technology, University of Oklahoma, Norman, OK 73019, USA}

\author{Charles M. Brooks}
\affiliation{Department of Physics, Harvard University, Cambridge, 02138 USA}

%%%%%%%%%%%%%%%%%%%%%

% Sample synthesis - LaNiO3,
\author{Hanjie Guo}
\affiliation{Max Planck Institute for Chemical Physics of Solids, N{\"o}thnitzer Str. 40, 01187 Dresden, Germany}
\author{Alexander C. Komarek}
\affiliation{Max Planck Institute for Chemical Physics of Solids, N{\"o}thnitzer Str. 40, 01187 Dresden, Germany}

%%%%%%%%%%%%%%%%%%%%%

% Sample synthesis - La2CuO4,

\author{Takanori Taniguchi}
\affiliation{Institute for Materials Research, Tohoku University, Katahira, Sendai 980-8577, Japan}
\author{Masaki Fujita}
\affiliation{Institute for Materials Research, Tohoku University, Katahira, Sendai 980-8577, Japan}

%%%%%%%%%%%%%%%%%%%%%%%

%Bosses

\author{Julia A. Mundy}
\affiliation{Department of Physics, Harvard University, Cambridge, 02138 USA}
\affiliation{John A. Paulson School of Engineering and Applied Sciences, Harvard University, Cambridge, 02138 USA}

\author{Takashi Mizokawa}
\affiliation{Department of Applied Physics, Waseda University, 3-4-1 Okubo, Shinjuku-ku, Tokyo 169-8555, Japan}

\author{Liu Hao Tjeng}
\affiliation{Max Planck Institute for Chemical Physics of Solids, N{\"o}thnitzer Str. 40, 01187 Dresden, Germany}

\author{Berit H. Goodge}
\affiliation{Max Planck Institute for Chemical Physics of Solids, N{\"o}thnitzer Str. 40, 01187 Dresden, Germany}

\author{Atsushi Hariki}
\affiliation{Department of Physics and Electronics, Osaka Metropolitan University, 1-1 Gakuen-cho, Nakaku, Sakai, Osaka 599-8531, Japan}

\date{\today}% It is always \today, today,

\begin{abstract}
We present a comparative Ni core level photoemission study of La$_3$Ni$_2$O$_7$, Nd$_3$Ni$_2$O$_7$, and LaNiO$_3$ using both the Ni $2p$ and the Ni $1s$. We address the challenges in analyzing the widely investigated Ni $2p$ spectra arising from the substantial overlap in energy of the Ni $2p$ with the La $3d$. We show that on the other hand the deep Ni $1s$ core level does provide a clean view on the intrinsic electronic excitations and we highlight its potential to resolve detailed differences in the electronic structure within the strongly correlated Ruddlesden–Popper series La$_{n+1}$Ni$_n$O$_{3n+1}$.
	
\end{abstract}

\maketitle
%%%%%%%%%%%%%%%%%%%%%%%%%%%%%%%%%%%
\section{Introduction}

Core-level x-ray photoelectron spectroscopy (PES) is a powerful probe for studying correlated materials. The x-ray excitation creates a core hole acting as a test charge to the electronic system and promoting dynamical charge responses, known as charge-transfer (CT) screening, from the surrounding environment~\cite{HuefnerBook,groot_kotani,Veenendaal1993,Hariki17}. This produces distinct features that contain information on the chemical bonding and the energetics of the various states involved. For 3$d$ transition metal oxides (TMOs), the 2$p$ core level has traditionally been the choice in PES~\cite{Ghijsen1988,Okada1989,Bocquet1992,Bocquet1992SFO,Saitoh1995}, as it has the sharpest spectral features among the core levels that can be reached with the commonly used Al~K$\alpha$ source, and it is routinely applied to novel TMOs to determine reliable parameters for modelling their electronic structure~\cite{Higashi21,Takegami22,Takegami2024_SFO,Takegami2025_NNO}, as well as to extract information about their low-energy physics including magnetic properties~\cite{Takegami23_LCO,Mondal2023,Suga2009,Eguchi2008,Horiba2004,Taguchi2005,Taguchi2005,Hariki2013}.

Recently, the discovery of superconductivity in the nickelates has generated tremendous research efforts from the condensed matter community~\cite{wang2024experimental,puphal2026superconductivity, Goodge2025}. In particular, La-based Ruddlesden--Popper nickelates La$_{n+1}$Ni$_{n}$O$_{3n+1}$ ($n = 2, 3$)~\cite{Sun2023,zhu2024superconductivity,ko2024signatures,zhou2025ambient,li2026bulk} highlight questions about the interplay between lattice and electronic structure, including the systematics of formal Ni valence states as well as the hybridization and charge transfer of Ni with the ligands~\cite{Luo2023, zhang2023electronic, Sakakibara24, lu2024interlayer, xia2025sensitive, rhodes2024structural, zhao2025electronic}.

The use of $2p$ core level PES for this important class of La-based nickelates is so far rather limited, however, due to energy overlap of the Ni $2p$ and La $3d$ core levels~\cite{Mickevicius2006,Yamagami2021,Takegami2024_LNO}. This makes analysis of the 2$p$ core-level spectrum difficult and can lead to different interpretations of the electronic structure of, for example, La$_3$Ni$_2$O$_7$~\cite{Takegami2024_LNO,Liu22}.

In this work, we investigate the electronic structure of La$_3$Ni$_2$O$_7$, Nd$_3$Ni$_2$O$_7$, and LaNiO$_3$ with two different spectroscopic methods. First, we explore to what extent the overlapping Ni $2p$ and La $3d$ spectra can be disentangled by using other (i.e., non-$3d$) La core levels as reference, and by using the La $3d$ from Ni-free La compounds. Second, we employ Ni $1s$ core level spectroscopy which offers some advantages. The multiplet interaction between the 1$s$ core hole and the 3$d$ electrons is negligibly weak, which allows a clearer identification of intrinsic CT excitations, as previously demonstrated for Cr- and V-containing  systems~\cite{Eguchi2008,Suga2009,Yamaguchi2024_1s,Woicik2025}. Furthermore, the absence of spin-orbit coupling (SOC) in the 1$s$ avoids overlap of features that occur in 2$p$ PES when the 2$p_{1/2}$ and 2$p_{3/2}$ separation is comparable or smaller than the CT excitations, as illustrated in several Ti and V systems~\cite{Woicik2015,Woicik2020,Hariki2022_1s}. Still, a key question is how the core-hole lifetime of the $1s$~\cite{Woicik2015,Miedema2015,Regoutz2018,Woicik2018,Ghiasi2019,Woicik2020} affects the quality of the information in comparison to that obtained from the $2p$ core level PES standard, which we consider here for the specific case of Ni.

We employ PES using hard x-rays (HAXPES) in order to reach the deep Ni $1s$ core level. HAXPES offers an additional advantage of sufficient bulk sensitivity to avoid concerns of surface-dominated signals~\cite{HAXPESbook2016,Kalha_2021_HAXPES_review}, which is particularly important for novel materials with challenging synthesis.

%%%%%%%%%%%%%%%%%%%%%%%%%%%%%%%%%%%%%%%%%%%%%%%%%%%%%%%%%%

\section{Methods}

%%%%%%%%%%%% synthesis %%%%%%%%%%%%%
Nd$_3$Ni$_2$O$_7$ (LaAlO$_3$) and La$_3$Ni$_2$O$_7$ (NdGaO$_3$) thin films were synthesized with reactive ozone-assisted molecular beam epitaxy following a similar procedure outlined in Refs.~\cite{pan2022synthesis, ferenc2023limits}; a modified shuttering sequence was used for the La$_3$Ni$_2$O$_7$ thin film, adapted from Ref.~\cite{Lee2014}. LaNiO$_3$ single crystals were grown in a 125~bar oxygen pressure using the floating zone technique in a high pressure mirror furnace (HKZ, Scidre), as described in~\cite{Guo2018}. Sizable La$_2$CuO$_4$ single crystals were grown by the standard floating-solvent traveling-zone method. The grown crystal was annealed in one bar of O$_2$ gas to minimize oxygen deficiencies.

%%%%%%%%%%%%%% Beamtime %%%%%%%%%%
HAXPES measurements were performed at the Max-Planck-NSRRC HAXPES end station with MB Scientific A-1 HE analyzer, Taiwan undulator beamline BL12XU of SPring-8~\cite{takegami2019}. Photon energies of $h\nu=6.5$~keV and 10~keV with resolutions of around 270~meV and 320~meV respectively were used. Measurements were performed at 80~K.

%%%%%%%%%%%%%% DMFT %%%%%%%%%%
To interpret the experimental HAXPES data, we simulate the Ni core-level spectra using density functional theory (DFT) combined with dynamical mean-field theory (DMFT). We adopt the DFT+DMFT valence-band electronic structure obtained for Nd$_3$Ni$_2$O$_7$ in Ref.~\onlinecite{Takegami2025_NNO} and calculate the core-level spectra using the DFT+DMFT hybridization density $V(\varepsilon)$ with the same parameters as in the previous work. The $V(\varepsilon)$ is systematically modified to examine the influence of the hybridization on the Ni core-level spectra, as described in Sec.~III.

\section{Results}

%%%%%%%%%%%%%%%%%%%%%%%%%%%%%%%%%%%%%%%%%%%%%%%%%%%%%%%%%%%%%%%%
%%%%%%%%%%%%%%%%%%%%%%%%%%%%%%%%%%%%%%%%%%%%%%%%%%%%%%%%%%%%%%%%
\begin{figure}[]
\begin{center}
\includegraphics[width=0.98\columnwidth]{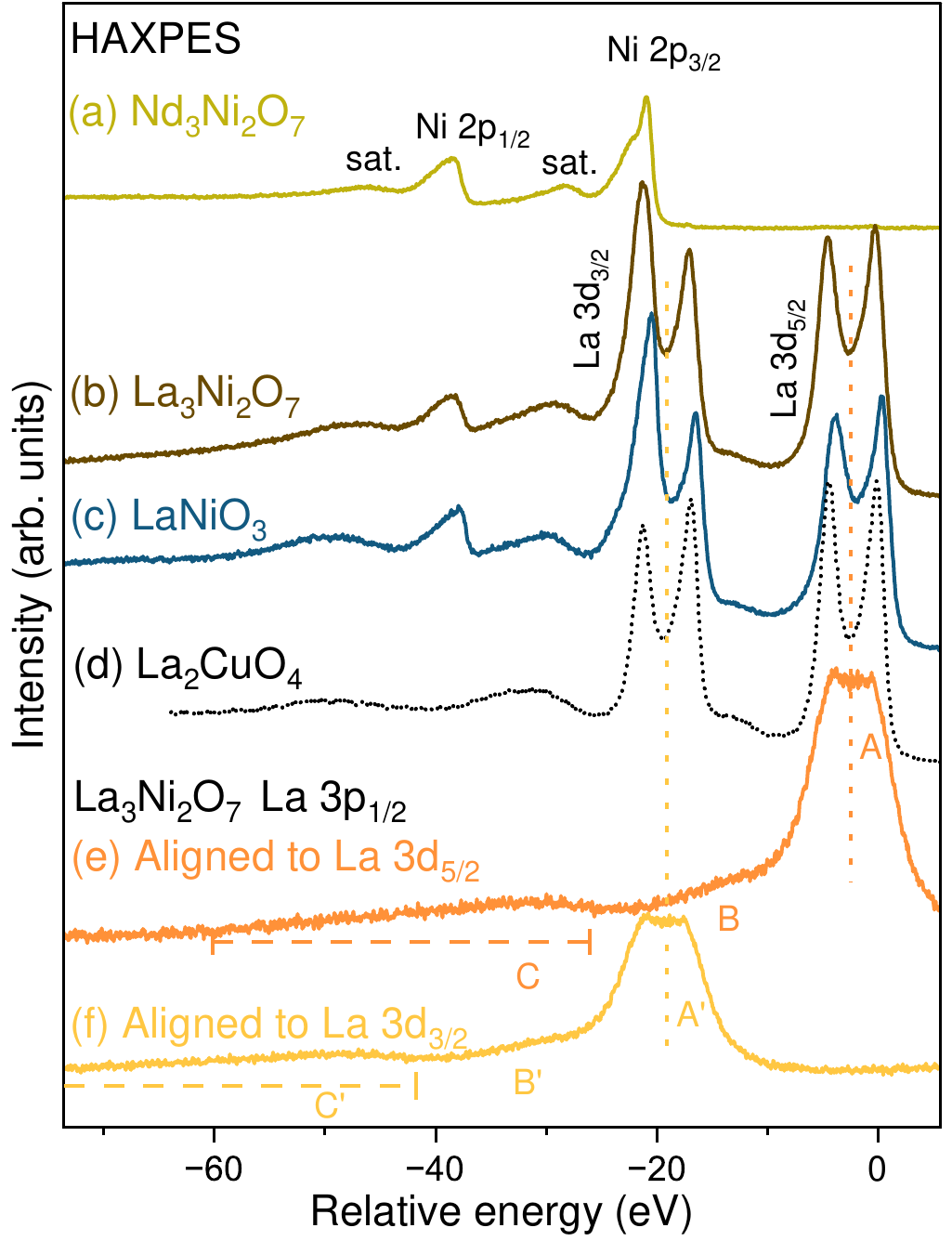}
\end{center}
\caption{
(a) Ni~$2p$ core level spectra of Nd$_3$Ni$_2$O$_7$~\cite{Takegami2025_NNO}. (b,c) Overlapping Ni~$2p$ and La~$3d$ core level spectra of (b) La$_3$Ni$_2$O$_7$ and (c) LaNiO$_3$. (d) La~$3d$ core level spectra of La$_2$CuO$_4$. (e,f) La~$3p_{1/2}$ spectra of La$_3$Ni$_2$O$_7$ shifted to match the centre of the (e) La~$3d_{5/2}$ and (f) La~$3d_{3/2}$ to illustrate the overlap of La-derived satellite features in the Ni~$2p$ region. The binding energies of the spectra have been shifted by (a-d) 833~eV, (e) 1205~eV and (f) 1222~eV in order to show the comparison in relative energy. 
}
\label{Fig_Cores}
\end{figure}
%%%%%%%%%%%%%%%%%%%%%%%%%%%%%%%%%%%%%%%%%%%%%%%%%%%%%%%%%%%%%%%%
%%%%%%%%%%%%%%%%%%%%%%%%%%%%%%%%%%%%%%%%%%%%%%%%%%%%%%%%%%%%%%%%

Figure~\ref{Fig_Cores} shows (a) the Ni~2$p$ spectra of Nd$_3$Ni$_2$O$_7$ from reference~\cite{Takegami2025_NNO}, together with  the Ni~2$p$ and La~3$d$ spectra of (b) La$_3$Ni$_2$O$_7$ and (c) LaNiO$_3$. Consistent with previous reports in the literature~\cite{Mickevicius2006, Yamagami2021,Takegami2024_LNO}, two sets of La~3$d$ double peaked core levels~\cite{Lam1980} are observed in the La-containing nickelates, with the La~$3d_{3/2}$ fully overlapping with the Ni~2$p_{3/2}$. Here we also observe that the La~$3d_{5/2}$ core level is noticeably different between materials, consistent with previous reports on the sensitivity of the La~$3d$ core level to differences in crystal structure and bonding environments~\cite{Lam1980,Kotani1992,Park1993}. This effectively prevents subtraction-based approaches using La spectra from other compounds to try and isolate the Ni peaks.

Concerning the Ni~2$p_{1/2}$ signal, it may appear at first glance that it is well separated from the La derived features. However, the La~3$d$ spectra of Ni-free La$_2$CuO$_4$ shown in Figure~\ref{Fig_Cores} (d) reveals the presence of high-energy satellites due to intrinsic and extrinsic energy loss excitations from the La~$3d$ in the same energy region of the Ni~2$p_{1/2}$ spectrum. This overlap significantly complicates interpretation of the Ni~2$p_{1/2}$ structures when comparing, for example, Nd- and La-based nickelate systems.

We can verify that such higher energy features are also present in the La core levels of La$_3$Ni$_2$O$_7$.
We show in Fig.~\ref{Fig_Cores}(e-f) the La~3$p_{1/2}$ spectra of La$_3$Ni$_2$O$_7$ shifted to match the main peak relative energies to those of La~$3d_{5/2}$ and $3d_{3/2}$. The La~3$p_{1/2}$ main peak (A,A') displays a double peak similar to that in La~3$d$, albeit with a significantly higher intrinsic broadening, as expected for $p$ core level spectra compared to their $d$ counterparts. At 10-15~eV higher energies, we observe a satellite-like feature (B,B'), followed by a very broad feature (C,C') extending 30-50~eV above the main peak, which might be derived from higher order excitations. These La features further hinder quantitative analysis of the Ni~2$p_{1/2}$ spectrum in La nickelates.

%%%%%%%%%%%%%%%%%%%%%%%%%%%%%%%%%%%%%%%%%%%%%%%%%%%%%%%%%%%%%%%%
%%%%%%%%%%%%%%%%%%%%%%%%%%%%%%%%%%%%%%%%%%%%%%%%%%%%%%%%%%%%%%%%
\begin{figure}[]
\begin{center}
\includegraphics[width=0.95\columnwidth]{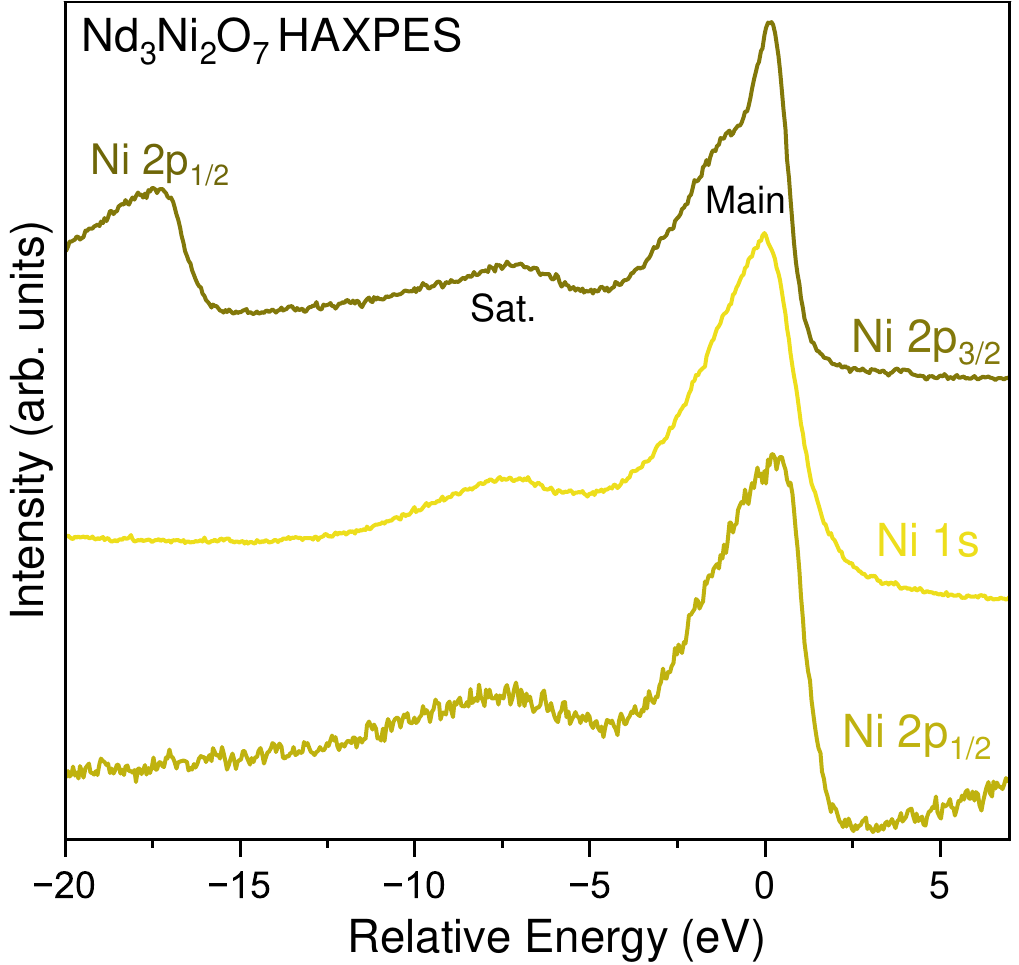}
\end{center}
\caption{
Comparison of the Ni~1$s$ core level spectra with the Ni~2$p_{3/2}$ and/or Ni~2$p_{1/2}$ for Nd$_3$Ni$_2$O$_7$. The binding energies of the spectra have been shifted by 8332.78~eV (Ni 1$s$) , 871.8~eV (Ni 2$p_{1/2}$), and 854.1~eV (Ni 2$p_{3/2}$).
}
\label{Fig_NdCores}
\end{figure}
%%%%%%%%%%%%%%%%%%%%%%%%%%%%%%%%%%%%%%%%%%%%%%%%%%%%%%%%%%%%%%%%
%%%%%%%%%%%%%%%%%%%%%%%%%%%%%%%%%%%%%%%%%%%%%%%%%%%%%%%%%%%%%%%

Because other approaches such as the use of photoionization cross section  differences~\cite{Weinen2015,takegami2019,Takegami22,Altendorf2023,Christovam2024} are not effective due to the nearly identical asymmetry parameters and similar photon energy dependencies of the Ni $2p$ and La $3d$ shells~\cite{trzhaskovskaya18}, we resort to another Ni shell that is free from energy overlap with non-Ni core levels.

Figure~\ref{Fig_NdCores} displays the Ni~1$s$ core-level spectrum together with the Ni~2$p_{3/2}$ and Ni~2$p_{1/2}$ spectra of the La-free compound Nd$_3$Ni$_2$O$_7$. These spectra enable a detailed and direct comparison of the intrinsic line shapes. All three exhibit strong similarities: a main asymmetric peak arising from non-local screening contributions~\cite{Takegami2025_NNO,Veenendaal1993,Hariki17} and a satellite about 7~eV above the main peak. Their energy splittings and intensity ratios are essentially identical across all core levels, indicating that the same underlying physics can be derived from each. The main distinction lies in the spectral broadening: the Ni~2$p_{3/2}$ spectrum is the sharpest, with local and non-local screening components clearly resolved. The Ni~1$s$ is somewhat broader than the 2$p_{3/2}$, but sharper than the 2$p_{1/2}$. Overall, we conclude that the Ni~1$s$ is a suitable alternative for the Ni~$2p$ core level spectroscopy.

%%%%%%%%%%%%%%%%%%%%%%%%%%%%%%%%%%%%%%%%%%%%%%%%%%%%%%%%%%%%%%%%

\begin{figure}[]
\begin{center}
\includegraphics[width=0.99\columnwidth]{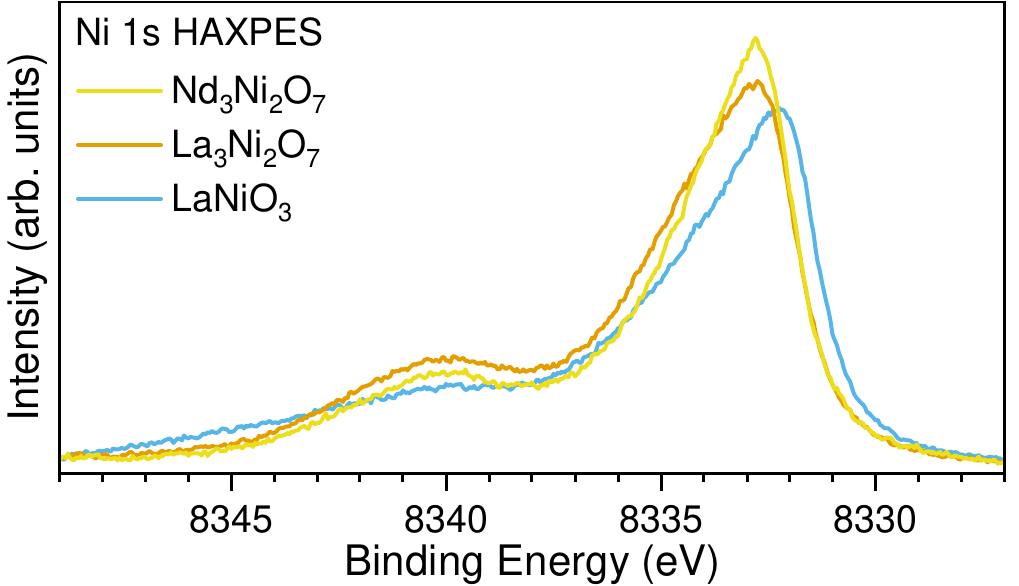}
\end{center}
\caption{Ni~1$s$ core-level spectra of Nd$_3$Ni$_2$O$_7$ (yellow line), La$_3$Ni$_2$O$_7$ (orange line), and LaNiO$_3$ (blue line).}
\label{Fig_comparison}
\end{figure}

%%%%%%%%%%%%%%%%%%%%%%%%%%%%%%%%%%%%%%%%%%%%%%%%%%%%%%%%%%%%%%%%

Figure~\ref{Fig_comparison} shows the Ni $1s$ spectra of Nd$_3$Ni$_2$O$_7$, La$_3$Ni$_2$O$_7$, and LaNiO$_3$.
While the overall line shapes and satellite structures are rather similar, the Ni $1s$ spectra do show material-dependent variations. First of all, we observe that there is a clear distinction between bilayer compounds Nd$_3$Ni$_2$O$_7$ and La$_3$Ni$_2$O$_7$ as compared to perovskite LaNiO$_3$. LaNiO$_3$ displays a much more asymmetric main peak at a lower energy, and a weaker, more extended satellite. While significant differences in the electronic structure between these two sets of compounds can be expected, it is worth noting that in the Ni~2$p$ spectra, as shown in Figs.~\ref{Fig_Cores} and \ref{Fig_NdCores}, the distinctions are far from evident. This demonstrates the clear advantage of using the Ni~1$s$ core level.

\begin{figure}[]
\begin{center}
\includegraphics[width=0.99\columnwidth]{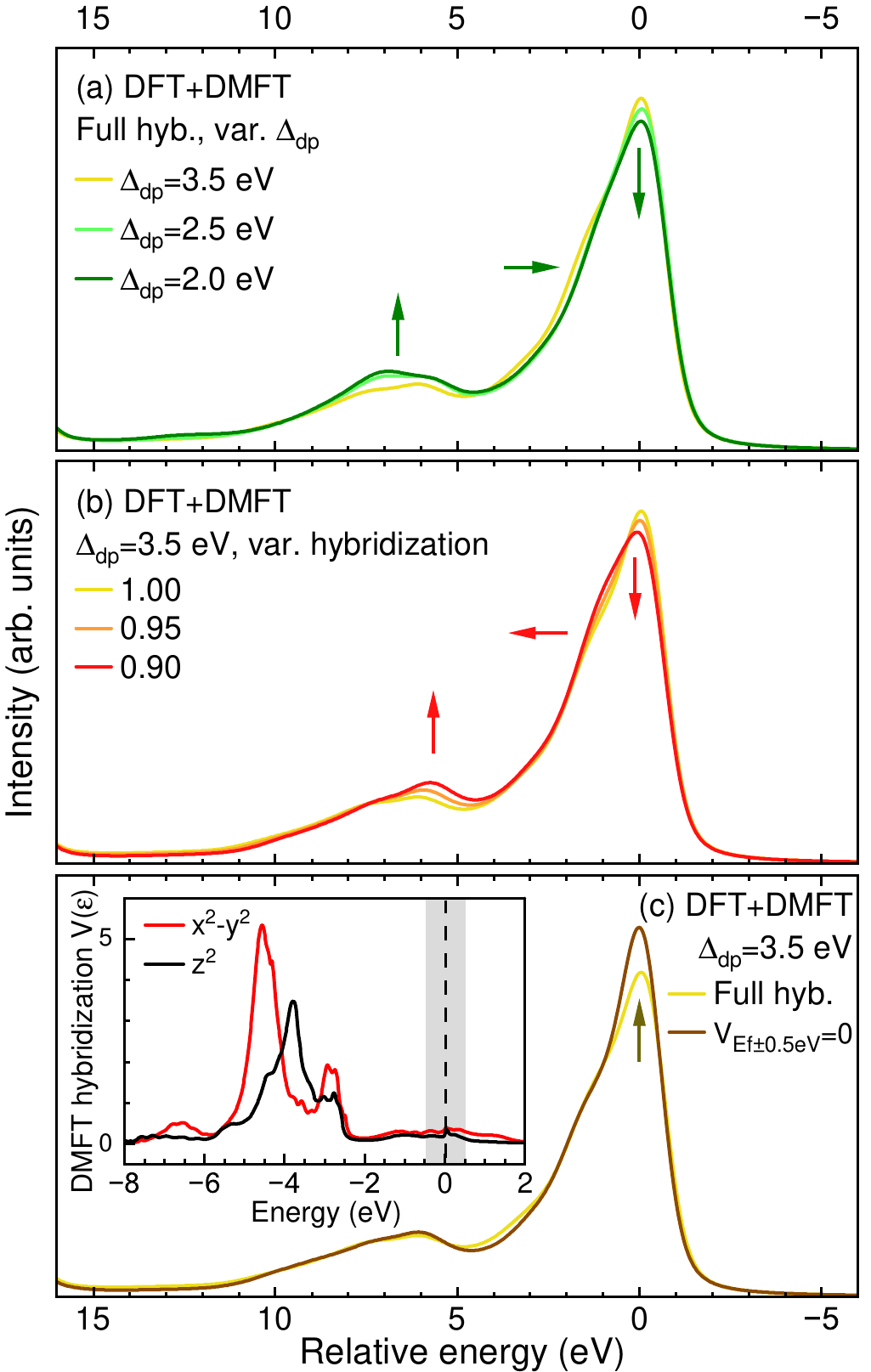}
\end{center}
\caption{(a) DFT+DMFT calculations of the Ni core-level (2$p$) spectra of Nd$_3$Ni$_2$O$_7$ for different CT energies $\Delta_{dp}$, as discussed in Ref.~\onlinecite{Takegami2025_NNO}. (b) Spectra calculated for $\Delta_{dp}=3.5$~eV with hybridization densities uniformly scaled over the entire energy range to 95\% (orange) and 90\% (red) of the original value (yellow). The original hybridization density $V(\varepsilon)$ for $\Delta_{dp}=3.5$~eV is shown in the inset of panel (c). The hybridization densities $V(\varepsilon)$ for different $\Delta_{dp}$ are provided in Ref.~\onlinecite{Takegami2025_NNO}.
(c) Spectra calculated using a modified hybridization density $V(\varepsilon)$ in which the hybridization amplitude within the energy window $E_F \pm 0.5$~eV (shaded area in the inset) is artificially set to zero. The original spectrum (yellow) with the full hybridization density $V(\varepsilon)$ is also shown for comparison.}
\label{Fig_VEf0}
\end{figure}

We now focus on more detailed comparison between La$_3$Ni$_2$O$_7$ and Nd$_3$Ni$_2$O$_7$. The La compound has a slightly broader main peak with reduced intensity, and the satellite spectral weight is enhanced compared to its Nd counterpart. To clarify the electronic structure changes underlying the observed spectral differences, we perform DFT+DMFT calculations, which were previously applied to simulate the Ni~2$p$ spectrum of Nd$_3$Ni$_2$O$_7$~\cite{Takegami2025_NNO}. Because the La and Nd samples are grown on different substrates and may also involve different levels of oxygen deficiency, a fully quantitative description is not feasible. We thus compare spectra computed with varying the CT energy $\Delta_{dp}$ in Fig.~\ref{Fig_VEf0}(a), taken from Ref.~\onlinecite{Takegami2025_NNO}, with supplemental spectra computed using hybridization densities $V(\varepsilon)$ rescaled from those optimized for Nd$_3$Ni$_2$O$_7$ in Fig.~\ref{Fig_VEf0}(b), to investigate trends in the $d$-level energies and hybridization in these nickelates.

As indicated by the arrows in Figs.~\ref{Fig_VEf0}(a) and \ref{Fig_VEf0}(b), reducing $\Delta_{dp}$ and the hybridization leads to similar overall changes in the spectra, namely an enhancement of the satellite spectral weight and a suppression of the low-binding-energy side of the main peak. A qualitative difference, however, appears in the width of the main line:~it becomes broader upon reducing the hybridization, whereas it becomes narrower upon reducing $\Delta_{dp}$. In Fig.~\ref{Fig_comparison}, La$_3$Ni$_2$O$_7$ indeed exhibits a broader main line compared to Nd$_3$Ni$_2$O$_7$, indicating that changes in hybridization provide the dominant contribution to the observed spectral differences between the two samples. Here we note that the Nd$_3$Ni$_2$O$_7$ and La$_3$Ni$_2$O$_7$ thin films are grown on LaAlO$_3$ and NdGaO$_3$ substrates, imposing compressive and tensile strain \cite{ferenc2023limits,bhatt2026structural}, respectively, and thus weaker hybridization is expected in La$_3$Ni$_2$O$_7$.

%%%%%%%%%%%%%%%%%%%%%%%%%%%%%%%%%%%%%%%%%%%%%%%%%%%%%%%%%%%%%%%%

%%%%%%%%%%%%%%%%%%%%%%%%%%%%%%%%%%%%%%%%%%%%%%%%%%%%%%%%%%%%%%%%
%%%%%%%%%%%%%%%%%%%%%%%%%%%%%%%%%%%%%%%%%%%%%%%%%%%%%%%%%%%%%%%%

We remark that the core-level spectrum is also expected to be sensitive to subtle changes in the metallic states around the Fermi energy $E_F$ through screening effects in the PES final states. To examine its impact, Fig.~\ref{Fig_VEf0}(c) shows a spectrum calculated using an artificial hybridization density in which the hybridization amplitude near $E_F$ is eliminated numerically. The corresponding hybridization densities $V(\varepsilon)$ are shown in the inset of Fig.~\ref{Fig_VEf0}(c). This substantially affects the intensity on the low-binding-energy side of the main line, which corresponds to final states associated with nonlocal screening~\cite{Takegami2025_NNO,Veenendaal1993,Hariki17}, while the other spectral features remain unchanged.

These results demonstrate the sensitivity of the core-level spectrum to various aspects of the electronic structure. On the one hand, this highlights the complexity of the spectrum and the challenges in obtaining a quantitative comparison between two compounds where several of these aspects might be changing at the same time. On the other hand, by performing a systematic study using samples grown under different conditions or on different substrates, as well as with varying numbers of layers ($n$) in the Ruddlesden--Popper nickelates R$_{n+1}$Ni$_n$O$_{3n+1}$, this sensitivity of the core level spectrum should allow for isolated determination of the specific effects these individual parameters have on nickelate electronic structure, laying a roadmap for detailed characterization.

\section{Conclusions}
We compared Ni 1$s$ and 2$p$ core-level hard x-ray photoelectron spectroscopy of the isostructural bilayer Ruddlesden–Popper nickelates La$_3$Ni$_2$O$_7$ and Nd$_3$Ni$_2$O$_7$, together with the reference compound LaNiO$_3$. We showed that a reliable extraction of the intrinsic Ni 2$p$ line shape, including its 2$p_{1/2}$ spin–orbit component, is not feasible for the La-based nickelates because of the overlap with the La 3$d$ core level and the presence of additional La-derived satellites. In contrast, we demonstrated that Ni 1$s$ core-level photoelectron spectroscopy is a suitable alternative and can resolve subtle differences in the Ni valence electronic structure among the nickelates. The spectral redistribution between the main line and the charge-transfer satellite observed for La$_3$Ni$_2$O$_7$ in comparison to Nd$_3$Ni$_2$O$_7$ can be interpreted using DFT+DMFT in terms of changes in the charge transfer energy and hybridization strength of Ni with surrounding ligands, which in turn can be associated with the effect of different tensile strain or chemical doping.

\section*{Acknowledgments}

D.T.~acknowledges the support by the Deutsche Forschungsgemeinschaft (DFG, German Research Foundation) under the Walter Benjamin Programme, Projektnummer 521584902. A.H.~was supported by JSPS KAKENHI Grant Numbers 25K00961, 25K07211, 23H03816, 23H03817, the 2025 Osaka Metropolitan University (OMU) Strategic Research Promotion Project (Young Researcher).
We acknowledge the support for the measurements from the Max Planck-POSTECH-Hsinchu Center for Complex Phase Materials. 
G.A.P. and D.F.S. are primarily supported by U.S. Department of Energy (DOE), Office of Basic Energy Sciences, Division of Materials Sciences and Engineering, under Award No. DE SC0021925; and by NSF Graduate Research Fellowship Grant No. DGE-1745303. G.A.P. acknowledges additional support from the Paul \& Daisy Soros Fellowship for New Americans. Q.S. was supported by the Science and Technology Center for Integrated Quantum Materials, NSF Grant No. DMR-1231319. J.A.M. acknowledges support from the U.S. Department of Energy (DOE), Office of Basic Energy Sciences, Division of Materials Sciences and Engineering, under Award No. DE SC0021925. Materials growth was supported by PARADIM under National Science Foundation (NSF) Cooperative Agreement No. DMR-2039380.

%\appendix

%\section{Appendix?}

\end{document}